\documentclass{PoS}

\usepackage{subfigure}
\usepackage{capt-of}
\usepackage{amsmath}
\usepackage{url}

\newcommand{\de}{\mbox{d}}
\newcommand{\GeV}{{\rm\ GeV}}

\title{Probing Dark Matter Long-lived Mediators with Solar $\gamma$ rays}

\ShortTitle{Probing Dark Matter Long-lived Mediators with Solar $\gamma$ rays}

\author{\speaker{Michele Lucente},$^{a}$\thanks{Supported by the Fonds de la Recherche Scientifique - FNRS under Grant No. IISN 4.4512.10.}\phantom{o} Chiara Arina,$^{a}$ Mihailo Backovi\'c,$^{a}$ and Jan Heisig$^{b}$
       \\
       \\
      \llap{$^{a}$}Center for Cosmology, Particle Physics and Phenomenology (CP3) \\ Universit\'e catholique de Louvain, B-1348 Louvain-la-Neuve, Belgium\\
      \llap{$^{b}$}Institute for Theoretical Particle Physics and Cosmology\\ RWTH Aachen University, D-52056 Aachen, Germany\\ \\
      E-mail: \email{michele.lucente@uclouvain.be}, \email{chiara.arina@uclouvain.be}, \email{mihailo.backovic@uclouvain.be}, \email{heisig@physik.rwth-aachen.de}}

\abstract{
We show that solar $\gamma$-ray observations can provide a complementary probe of Dark Matter in scenarios where the interactions with the Standard Model proceed via long-lived mediators. 
For illustration we consider a simplified model which provides solar $\gamma$-ray
fluxes observable with the next generation $\gamma$-ray telescopes, while
complying with the existing experimental constraints.
Our results suggest that solar $\gamma$-ray fluxes can be orders of magnitude larger than the ones from the Galactic center, while being subject to low backgrounds. 
}

\FullConference{The European Physical Society Conference on High Energy Physics\\
		 5-12 July\\
		 Venice, Italy}

\begin{document}

\section{Introduction}

The majority of indirect Dark Matter (DM) searches with $\gamma$-ray telescopes target the Galactic center (GC) or dwarf spheroidal 
galaxies, exploiting the high DM density or the large mass-to-light ratio, respectively. However, other celestial targets can provide complementary or competitive information. The Sun, in particular, can serve as a nearby reservoir of DM~\cite{capture}: provided a non-vanishing DM-quark interaction is present, scattering off solar matter results in DM energy losses, leading to gravitational capture and DM accumulation in the center of the Sun, where it can annihilate. At higher orders in perturbation theory the same interaction will naturally couple DM to photons, enabling the production of $\gamma$ rays from DM annihilation. If the DM annihilation proceeds via intermediate states (mediators) long-lived enough so that they can escape from the Sun surface, it is possible to observe the resulting $\gamma$ rays on Earth. 

There are several reasons why the study of high-energy solar $\gamma$ rays can provide a powerful tool to probe DM models characterised by long-lived mediators~\cite{Arina:2017sng}: the Sun is a poor source of photons above the GeV energy scale, and uncertainties due to the DM density distribution are largely reduced with respect to GC observations. In addition, due to the relative proximity to Earth, solar $\gamma$-ray fluxes are potentially much larger than those originating from dwarfs or from the GC.

\section{Simplified model with Yukawa-like mediator-quark couplings}

In order to illustrate how solar $\gamma$ rays can be a competitive probe for DM models featuring long-lived mediators, we consider a simplified model with a fermionic DM candidate $X$ and a mixed scalar-pseudoscalar mediator $Y$, that couples to the Standard Model quarks $q_j$ via Yukawa-like couplings. The new interactions are described by the Lagrangian
\begin{eqnarray}
	\mathcal{L} & = & g_q y_{q_j} \, \bar{q_j} \left[ {\rm cos} \,\alpha +  i\, {\rm sin}\, \alpha\, \gamma_5 \right] q_j \,Y +  g_X \, \bar{X} \left[ {\rm cos} \,\alpha + i \, {\rm sin}\, \alpha\, \gamma_5 \right] X\, Y\,, \label{eq:lagr}
\end{eqnarray}
where $g_q$ and $g_X$ are coupling constants, $\alpha$ is the scalar-pseudoscalar mixing angle and $y_{q_j}\equiv \sqrt{2} m_{q_j} / v_h$ is the quark Yukawa coupling, with $v_h=246 \GeV$ and $m_{q_j}$ the quark mass. In this model, for $m_Y<m_X$ two DM particles $X$ can annihilate into a pair of mediators $Y$, each of them subsequently decays into a pair of photons via loop-induced diagrams. If $m_Y<2 m_\pi$, with $m_Y$ and $m_\pi$ the mediator and pion masses, respectively, only the decay channel into a pair of photons is open, since gluons would not be kinematically allowed to hadronise. This results in a box-shaped photon energy spectrum~\cite{Ibarra:2012dw}, \emph{i.e.}~$\de N_\gamma/\de E_\gamma \propto \Theta(m_X-E_\gamma) /m_X$ for $m_Y \ll m_X$.

\section{Parameter space and constraints}

\begin{figure}[htb]
\centering
\subfigure{
     \includegraphics[width=0.4\textwidth]{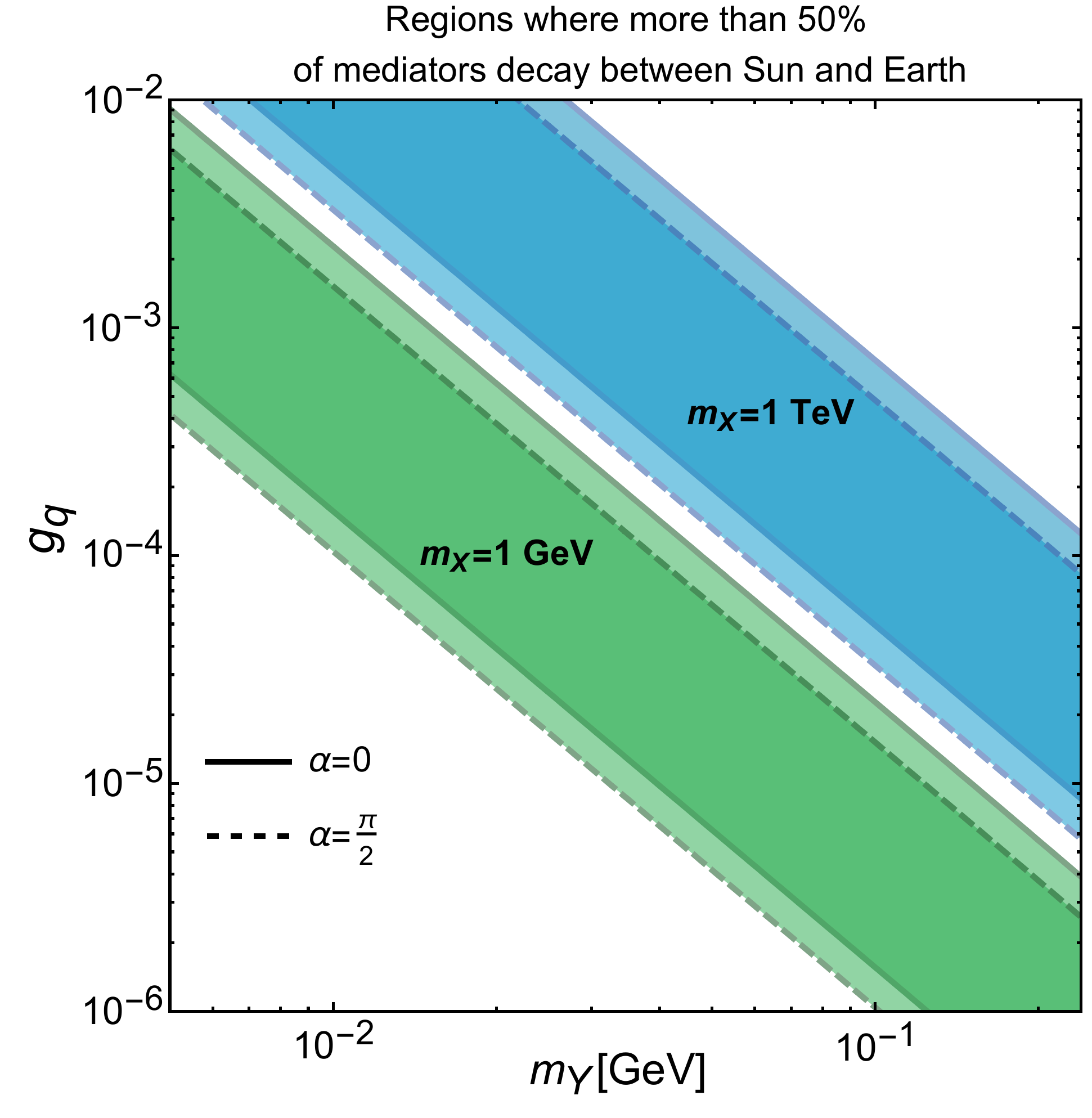}}
\hspace{0.1\textwidth}
\subfigure{\includegraphics[width=0.39\textwidth]{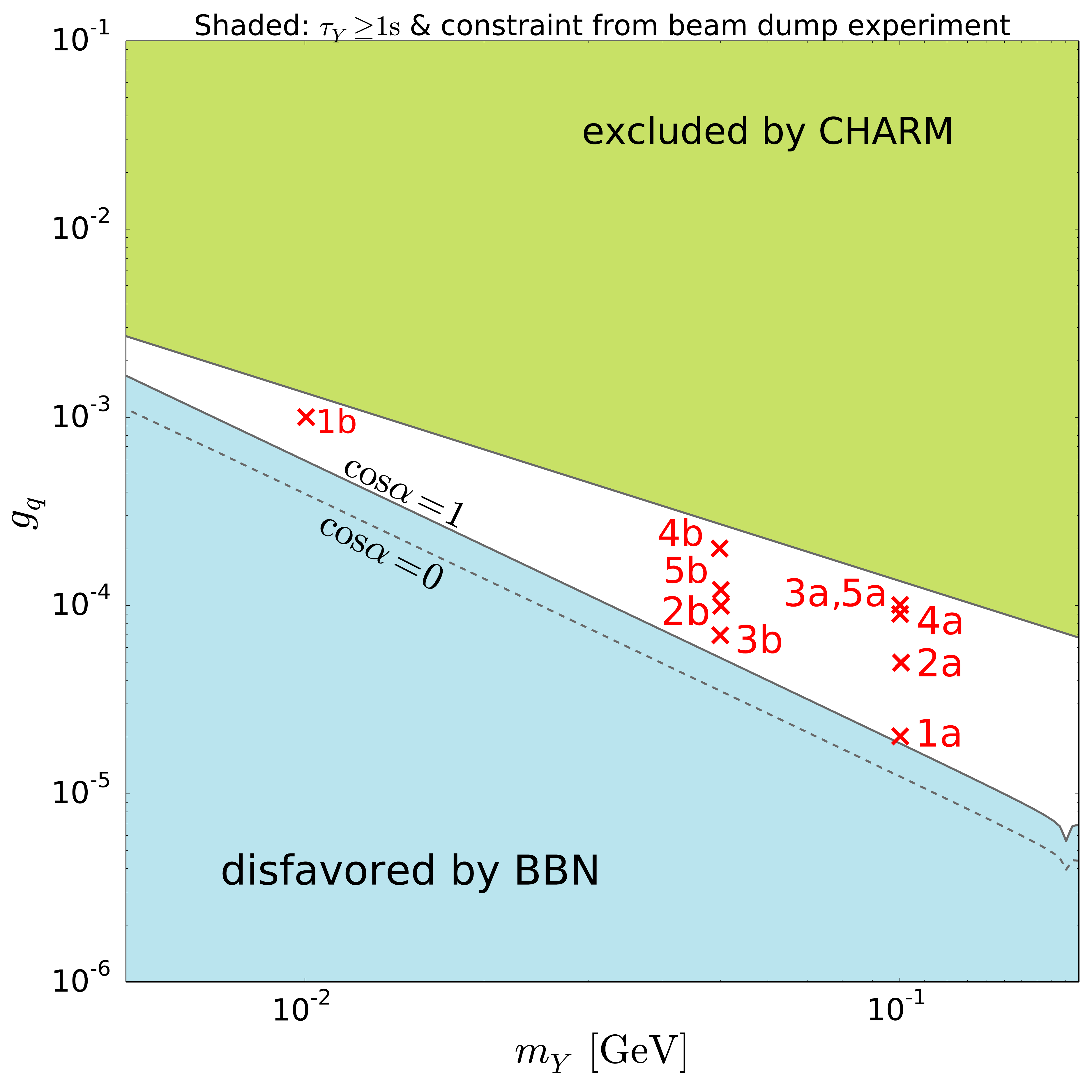}}
\caption{
\emph{Left}: Region of parameter space featuring at least 50\% of mediators decaying between the Sun surface and the Earth orbit, for two different values of $m_X$ and $\cos\alpha$.
\emph{Right}: Regions where the mediator lifetime exceeds 1s (blue) and excluded by the CHARM experiment~\cite{Bergsma:1985qz} (green).
}
 \label{Fig:par_space}
\end{figure}

For a mediator produced in the center of the Sun, the condition to escape the solar surface reads
$\tau_Y\left(m_X/m_Y\right) \gtrsim R_{\odot} $,
where $\tau_Y$ is the mediator lifetime, $m_X/m_Y$ its boost in the lab frame and $R_{\odot}$ the solar radius. Thus a relatively long mediator lifetime and/or a large hierarchy of masses between DM and mediator are necessary conditions in order to observe a $\gamma$-ray flux on Earth.
On the other hand, the decay length should not significantly exceed the Sun distance $D_\odot$. 
We define ${\cal F}_\text{det}$ as the fraction of mediators decaying between the Sun surface and the Earth orbit: 
the parameter space featuring ${\cal F}_\text{det}> 50\%$ is shown in Fig.~\ref{Fig:par_space} (left)
spanning a wide range in $g_q$.

The dominant DM annihilation channels in the early Universe are the $t$-channel process $X\bar{X} \rightarrow YY$ and, if $m_X > m_t$, the $s$-channel process $X\bar{X}\rightarrow t\bar{t}$.
Since the escape of the mediator from the Sun requires small $g_q$, the annihilation channel $X\bar{X}\rightarrow t\bar{t}$ is typically subdominant at freeze-out, leaving $X\bar{X} \rightarrow YY$ as the dominant process to fix the DM relic density. This leads to the requirement $g_X \approx 0.8 \sqrt{m_X/100\,\mathrm{GeV}}+\mathcal{O}(\cos^2\alpha)$~\cite{Arina:2017sng}.

Further constraints on the model arise from beam dump experiments and
Big Bang Nucleosynthesis (BBN). The right panel of Fig.~\ref{Fig:par_space} shows the 
allowed parameter space in the $m_Y$-$g_q$ plane, cornered from above by 
limits from CHARM~\cite{Arina:2017sng,Bergsma:1985qz}
and from below by BBN constraints, \emph{i.e.}~the conservative requirement $\tau_Y<1\,$s.
Notice that there is a significant overlap between the experimentally allowed parameter
space, shown in the right panel of Fig.~\ref{Fig:par_space}, and the one that provides a sizeable fraction of
mediator decays between the Sun and Earth, shown in the left panel.

Direct detection experiments impose constraints on the DM-nucleon scattering cross section.
The dominant contribution is the spin-independent part, $\sigma^{\rm SI}_{Xn} \propto g_q^2 g_Y^2 \cos^4\alpha$~\cite{Arina:2017sng}, which is not velocity suppressed.
In order to pass current direct detection constraints~\cite{Akerib:2016vxi}
very small values of $\cos\alpha$ (\emph{i.e.}~a dominant pseudoscalar component for the mediator) 
are required.

For our study we define a set of 10 benchmark model points. Their parameters are summarised in 
Table~\ref{Tab:benchmarks}. They are chosen such as to pass all the considered constraints. Furthermore, they
are in accordance with $\gamma$-ray limits from dwarf spheroidal galaxies~\cite{Arina:2017sng,indirect}.

\begin{table}[htb]
\centering
\scriptsize
\begin{tabular}{ccccccccc}\hline
Benchmark & $m_X [{\rm GeV}]$ & $m_Y [{\rm GeV}]$ & $g_X$ & $g_q$ & cos $\alpha$ &   $\tau_Y\,[\text{s}]$ &  ${\cal F}_\text{det}$  &   $\Phi_\gamma^{\odot} [{\rm cm}^{-2} \,\rm s^{-1}]$\\
\hline
 1a &  $10$ & $0.1$ & $0.24$ & $2 \times 10^{-5}$ &   
$0.01$ & $0.19$ & $0.88$ & $1.6 \times 10^{-15}$ \\
  1b &  $10$ & $0.01$ & $0.24$ & $0.001$ & $0.001$ & 
 $0.076$ & $0.97$ & $1.1 \times 10^{-10}$\\
 2a &  $100$ & $0.1$ & $0.76$ & $5 \times 10^{-5}$ & $0.012$ &
 $0.031$ & $0.93$ & $7 \times 10^{-11}$ \\
 2b &  $100$ & $0.05$ & $0.76$ & $0.0001$ & $0.004$ & 
$0.061$ & $0.96$ & $5.2 \times 10^{-12}$\\
 3a &  $300$ & $0.1$ & $1.4$ & $0.0001$ & $0.01$ & 
 $0.0076$ & $0.9$ & $5.7 \times 10^{-10}$\\
 3b &  $300$ & $0.05$ & $1.4$ & $7 \times 10^{-5}$ & $0.004$ &
 $0.12$ & $0.48$ & $2.1 \times 10^{-12}$\\
 4a &  $1000$ & $0.1$ & $2.5$ & $9 \times 10^{-5}$ & $0.011$ &
 $0.0094$ & $0.97$ & $1.3 \times 10^{-9}$ \\
 4b &  $1000$ & $0.05$ & $2.5$ & $0.0002$ & $0.003$  & 
 $0.015$ & $0.8$ & $2.2 \times 10^{-11}$\\
 5a &  $1800$ & $0.1$ & $3.4$ & $0.0001$ & $0.011$  & 
 $0.0076$ & $0.96$ & $1.4 \times 10^{-9}$\\
 5b &  $1800$ & $0.05$ & $3.4$ & $0.00012$ & $0.003$ & 
 $0.042$ & $0.28$ & $9 \times 10^{-13}$\\
\hline
\end{tabular}
\caption{Definition of benchmark model points, their mediator lifetimes, fractions of detectable decays and $\gamma$-ray fluxes.}
\label{Tab:benchmarks}
\end{table}

\section{Solar $\gamma$-ray fluxes}

\begin{figure}[htb]
\centering
\raisebox{0.005\textwidth}{\subfigure{\includegraphics[width=0.4\textwidth]{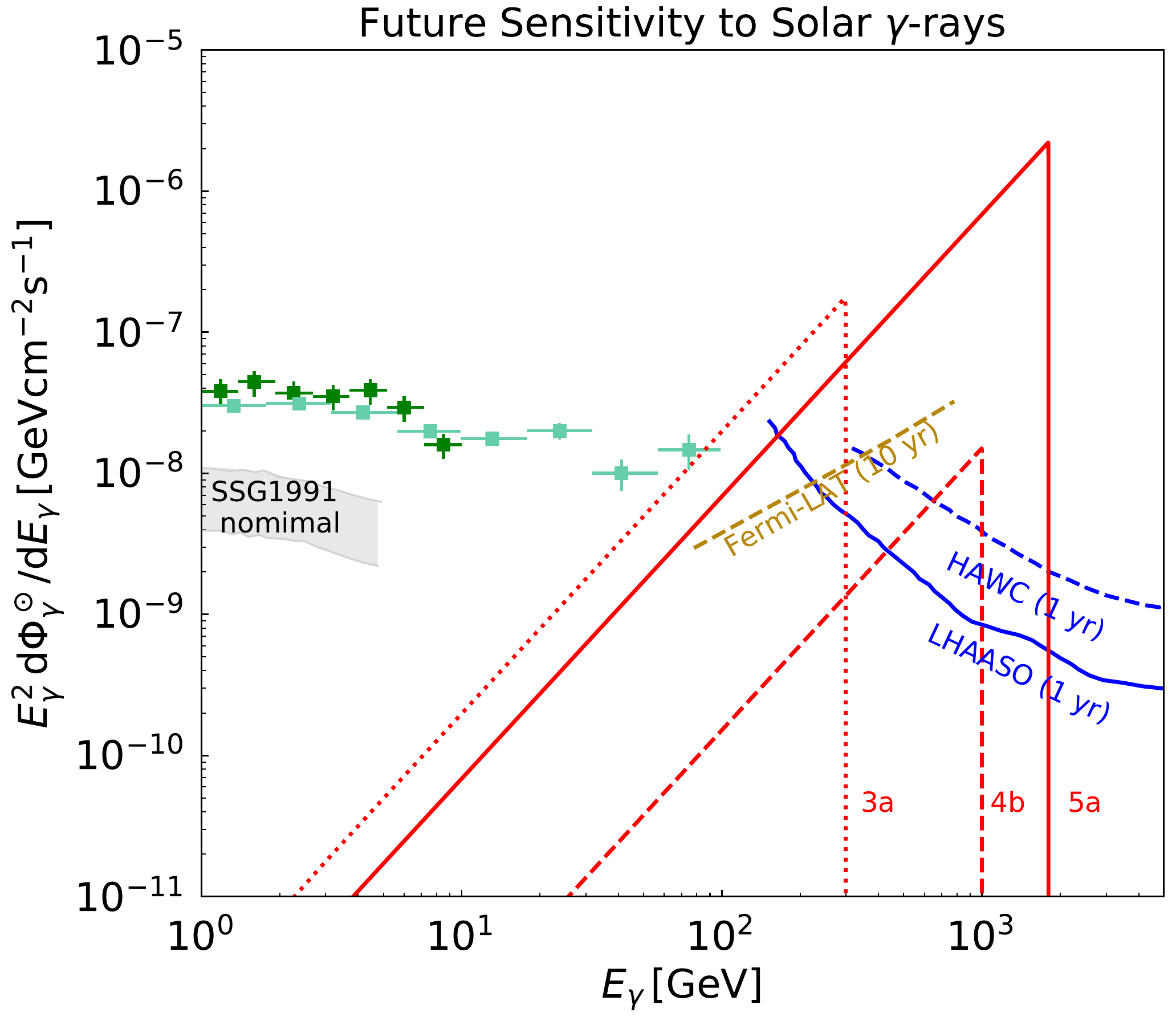}}}
\hspace{0.1\textwidth}
\subfigure{\includegraphics[width=0.37\textwidth]{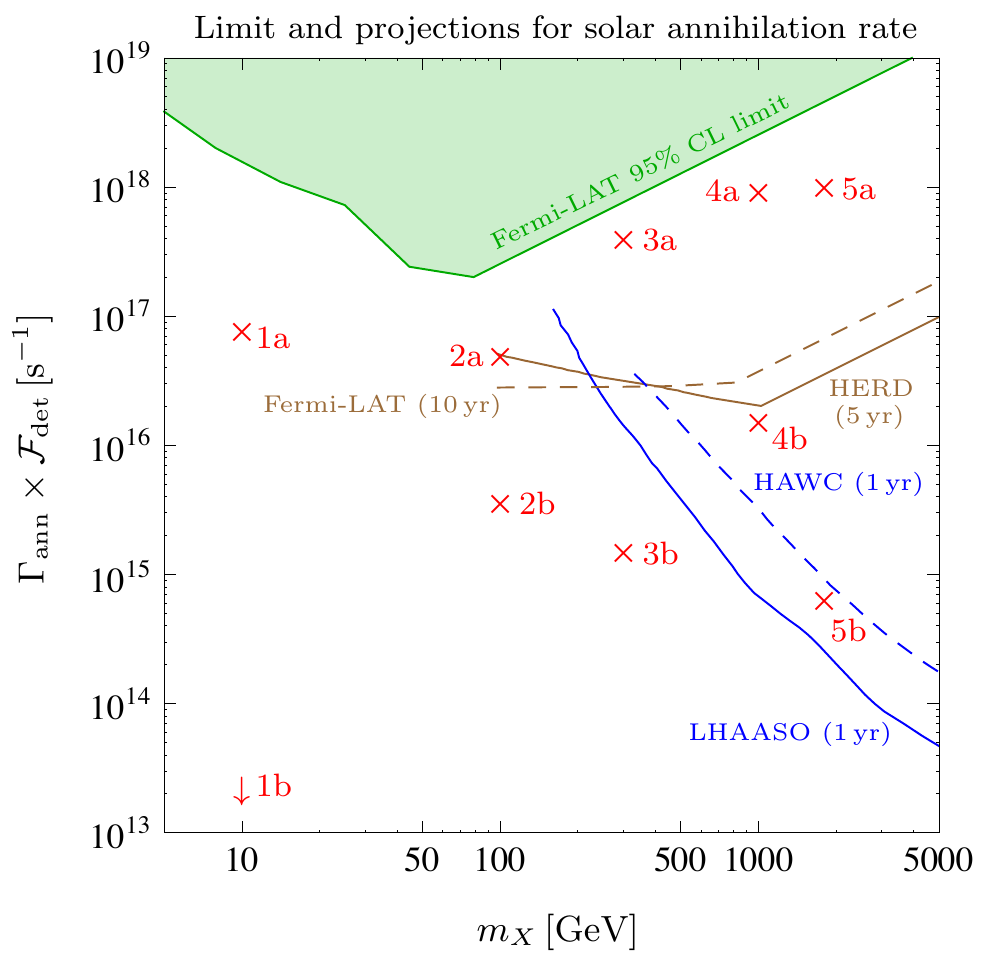}}
\caption{
\emph{Left}: Expected spectra from selected benchmark points and projected sensitivities of Fermi-LAT, HAWC and LHAASO~\cite{sensitivity}. The green points and the grey band correspond to the observed and expected solar $\gamma$-ray background~\cite{background}, respectively. \emph{Right}: Existing~\cite{background} and projected~\cite{sensitivity} exclusion limits on DM annihilation rate from solar $\gamma$-ray observations.
}
 \label{Fig:res}
\end{figure}

The evolution of the 
number $N$ of DM particles in the Sun is determined by the differential equation
\begin{equation}
\frac{{\rm d}N}{{\rm d}t} 
= C_\text{cap} - 2 \Gamma_{\rm ann },
\label{eq:Riccati}
\end{equation}
where $C_\text{cap}\propto\sigma^{\rm SI}_{Xn}$ and
$\Gamma_{\rm ann } \propto N^2 \langle \sigma v \rangle$ are the capture and 
annihilation rates, respectively. Assuming accumulation of DM during the entire Sun lifetime,
Eq.~\eqref{eq:Riccati} can be solved for $N$ and, hence, for $\Gamma_{\rm ann }$.\footnote{Note that
equilibrium between capture and annihilation (\emph{i.e.}~${\rm d}N/{\rm d}t = 0$) is not necessarily required. In fact, most benchmark points
do not reach equilibrium~\cite{{Arina:2017sng}}.}
For $m_Y\ll m_X$ the resulting differential flux is 
\begin{small}
$
\de \Phi_\gamma^\odot/\de E_\gamma \approx 4 \Theta(m_X-E_\gamma)  {\cal F}_\text{det}  \Gamma_{\rm ann}
/\left(4 \pi D_\odot^2 m_X\right)
$.
\end{small}
Despite the fact that the capture rate is strongly constrained by current direct detection experiments,
most of the resulting $\gamma$-ray fluxes (summarised in Table~\ref{Tab:benchmarks}) 
are within the reach of future $\gamma$-ray experiments. The right panel of Fig.~\ref{Fig:res}
compares the predicted effective annihilation rate $\Gamma_{\rm ann } {\cal F}_\text{det}$
to the corresponding projected sensitivities for Fermi-LAT, HERD, HAWC and LHAASO~\cite{sensitivity}.
The left panel of Fig.~\ref{Fig:res} exemplarily shows the spectra for the benchmark points $3a$, $4b$ and $5a$,
as well as the current measurements~\cite{background} and projected differential sensitivities. All points with $m_X>100\,$GeV lie outside the sensitivity of upcoming indirect 
detection experiments targeting dwarfs or the GC~\cite{indirectproj}, with solar $\gamma$ rays hence providing a superior sensitivity~\cite{Arina:2017sng}.

\section{Conclusion}

The Sun is a potential nearby reservoir of DM and a poor source of high-energy $\gamma$-ray backgrounds. Considering a simplified DM model, featuring a long-lived scalar-pseudoscalar mediator, we demonstrated the potential of solar $\gamma$-ray observations to probe DM annihilation with comparable or superior sensitivity with respect to observations in dwarfs and the Galactic center. In the considered model DM capture and annihilation proceed typically out of equilibrium, a general feature due to the strong bounds on DM-nucleon scattering from direct searches: provided an independent  measurement of the DM-nucleon cross section by future direct searches, solar $\gamma$ rays potentially allow for a determination of the DM annihilation cross section. 
Finally, solar $\gamma$-ray signals would point towards the existence of a long-lived mediator, a piece of information which could not be inferred from indirect detection in the GC/dwarf galaxies or in direct detection experiments.

\end{document}